\pdfoutput=1

\documentclass[review,11pt]{elsarticle}
\usepackage[margin=2.5cm]{geometry}

\usepackage{graphicx}
\usepackage{epstopdf}
\usepackage{caption}
\usepackage{subfig}

\usepackage{amssymb}
\usepackage{array}
\newcolumntype{C}[1]{>{\centering\let\newline\\\arraybackslash\hspace{0pt}}m{#1}}

\usepackage{amsthm}
\usepackage{amsmath}
\theoremstyle{definition}
\newtheorem{defn}{Definition}
\usepackage{algorithm}
\usepackage{algorithmic}

\biboptions{authoryear}

\usepackage[pscoord]{eso-pic}
\usepackage{textcomp}
\usepackage{hyperref}
\newcommand{\placetextbox}[3]{
  \setbox0=\hbox{#3}
  \AddToShipoutPictureFG*{
    \put(\LenToUnit{#1\paperwidth},\LenToUnit{#2\paperheight}){\vtop{{\null}\makebox[0pt][c]{#3}}}%
  }%
}%
\renewcommand*{\today}{February 9, 2019}

\journal{Expert Systems with Applications}

\begin{document}

\placetextbox{0.5}{0.999}{\normalfont \textcopyright 2020. This manuscript version is made available under the CC-BY-NC-ND 4.0 license}

\placetextbox{0.5}{0.985}{\normalfont \url{http://creativecommons.org/licenses/by-nc-nd/4.0/}}

\placetextbox{0.5}{0.965}{\normalfont Author accepted manuscript, published in ``Expert Systems With Applications, 2019, 131: 60--70''.}

\placetextbox{0.5}{0.950}{\normalfont DOI: \url{https://doi.org/10.1016/j.eswa.2019.04.041}.}%

\begin{frontmatter}

\title{Application of multi-objective optimization to blind source separation}

\author[a]{Guilherme Dean Pelegrina\corref{mycorrespondingauthor}}
\cortext[mycorrespondingauthor]{Corresponding author. Phone number: +55 19 982627361}
\ead{pelegrina@decom.fee.unicamp.br}

\author[b]{Romis Attux}
\ead{attux@dca.fee.unicamp.br}

\author[a]{Leonardo Tomazeli Duarte}
\ead{leonardo.duarte@fca.unicamp.br}

\address[a]{School of Applied Sciences, University of Campinas, 1300 Pedro Zaccaria Street, 13484-350 Limeira, Brazil}
\address[b]{School of Electrical and Computer Engineering, University of Campinas, 400 Albert Einstein Avenue, 13083-852 Campinas, Brazil}

\begin{abstract}
Several problems in signal processing are addressed by expert systems which take into account a set of priors on the sought signals and systems. For instance, blind source separation is often tackled by means of a mono-objective formulation which relies on a separation criterion associated with a given property of the sought signals (sources). However, in many practical situations, there are more than one property to be exploited and, as a consequence, a set of separation criteria may be used to recover the original signals. In this context, this paper addresses the separation problem by means of an approach based on multi-objective optimization. Differently from the existing methods, which provide only one estimate for the original signals, our proposal leads to a set of solutions that can be utilized by the system user to take his/her decision. Results obtained through numerical experiments over a set of biomedical signals highlight the viability of the proposed approach, which provides estimations closer to the mean squared error solutions compared to the ones achieved via a mono-objective formulation. Moreover, since our proposal is quite general, this work also contributes to encourage future researches to develop expert systems that exploit the multi-objective formulation in different source separation problems.
\end{abstract}


\end{frontmatter}

\section{Introduction}
\label{sec:intro}

In signal processing, a large number of problems are addressed by means of an optimization model whose cost function is associated with a single characteristic of the considered signal. However, in several practical situations, a set of characteristics is available for the system user. Even if a wide variety of information on the sources is available, most of existing methods do not take into account this set of information. Therefore, an effort should be conducted in order to exploit more information so the resulting signal processing systems is able to perform well even in difficult scenarios.

A typical situation that can be configured in this scenario is the Blind Source Separation (BSS) problem~\citep{Comon2010}. Introduced by H\'{e}rault, Jutten and Ans~\citeyearpar{Herault1985}, BSS problems consist in recovering a set of sources from the observation of a set of mixtures of these sources, without the knowledge of both original signals and mixing process. If we consider the particular case in which the aim is to recover a single source from the set of mixtures, the problem is called Blind Source Extraction (BSE)~\citep{Cichocki2002}. BSE is closely related to BSS, since the latter can be achieved by combining successive BSE techniques and, for instance, a deflation strategy~\citep{Hyvarinen2001}. Several practical situations are formulated as BSS or BSE problems, e.g. in audio signal processing~\citep{Choi1997,Xue2009}, chemical data analysis~\citep{Duarte2014a,Duarte2014b}, geophysical signal processing~\citep{Takahata2012,Batany2016} and manufacturing process analysis~\citep{Gandini2011}. Recent works also applied BSS techniques in face recognition problems~\citep{Bhowmik2019}. Moreover, in biomedical signal processing, one may find many works that deal with electroencephalogram (EEG)~\citep{Hsu2012,Charbonnier2016,Lafuente2017} and electrocardiogram (ECG)~\citep{Barros2001,Zhang2006,Shi2007,Yu2007,Yu2008,Yu2009} through BSS and BSE techniques.

BSS is generally carried out by formulating a separation criterion related to a single property of the sources, e.g., non-Gaussianity~\citep{Hyvarinen2001}, temporality~\citep{Barros2001} and sparsity~\citep{Nadalin2010}. This separation criterion leads to a mono-objective optimization problem whose solution provides the estimations of the source signals. However, in many applications, one does not find a single separation criterion that guarantees a perfect separation of the sources. In such situations, one may have more than one property to be exploited and, therefore, more than one separation criterion may be derived to deal with the BSS or BSE problem. For instance, Li, Liu and Principe~\citeyearpar{Li2007} used a correntropy measure in order to exploit both independence among sources and their temporal diversity. Shi and Zhang~\citeyearpar{Shi2007} considered temporality and non-Gaussianity properties in fetal electrocardiogram extraction. Duarte, Moussaoui and Jutten~\citeyearpar{Duarte2014b} discuss non-negativity and smoothness in chemical signals. Finally, Boukouvalas et al.~\citeyearpar{Boukouvalas2017} proposed an approach based on both sparsity and statistical independence to deal with functional magnetic resonance imaging.

A key aspect in the aforementioned works is that the signal properties are combined into a single criterion, leading to a mono-objective formulation. In the present paper, we consider a different approach to deal with BSS problems, which is based on multi-objective optimization. In this formulation, different optimization criteria, associated with different signal properties, are simultaneously optimized. The price to be paid is that, since one aims at optimizing more than two cost functions simultaneously, the optimality condition should take into account more than two values instead of only one, which is the case of the mono-objective formulation. Therefore, more sophisticated methods should be used to address this problem, such as the one based on evolutionary algorithms~\citep{Deb2001}.

However, in an expert system point of view, instead of a single solution for the addressed problem, one achieves a set of optimal solutions. Based on this set, also known as non-dominated set~\citep{Miettinen1999,Deb2001}, the system user can select the best solution according to his/her knowledge on the problem. Therefore, differently from the existing mono-objective approaches, a benefit of the proposed system is that it provides more information for the user to support his/her decision. It is worth noting that such expert system is suitable, for example, in biomedical~\citep{Vigario2008} and geophysical signal processing~\citep{Takahata2012}, in which the final decision is made by physicians and geophysicists, respectively.

In the literature, one may find works that consider the multi-objective optimization to deal with signal processing problems in specific situations. For instance, Wu et al.~\citeyearpar{Wu2018} applied this approach in low rank and sparse matrix decomposition. In the context of BSS, Phlypo et al.~\citeyearpar{Phlypo2006} and Goh et al.~\citeyearpar{Goh2016} consider the multi-objective optimization in electroencephalogram (EEG) signal processing. In a recent work, Pelegrina and Duarte~\citeyearpar{Pelegrina2018} also used this formulation combined with a memetic strategy to deal with a post-nonlinear mixture in chemical signals. However, since this formulation is not restricted to EEG or chemical signals, we propose a general expert system based on an evolutionary multi-objective algorithm which may be applied on any BSS problem. 

It is worth noting that this paper extends the technical contents that were partially presented in~\citep{Pelegrina2016}, which contain some initial results in blind source extraction. Moreover, this full version provides a discussion on the task of selecting a non-dominated solution. This procedure, as demonstrated in the numerical experiments, may lead to estimates closer to the mean squared error (MSE) solutions\footnote{We refer to the MSE solution as the estimates derived by a supervised approach that requires the knowledge of the original sources, which are unknown in real situations. We detail this procedure in Section~\ref{subsec:explin}.} compared to the mono-objective optimization of single criteria.

The rest of this paper is organized as follows. In Section~\ref{sec:theory}, we discuss the theoretical aspects of BSS problems and multi-objective optimization. Section~\ref{sec:mooapr} describes the proposed multi-objective approach. In Section~\ref{sec:exp}, we present the numerical experiments performed in this work and the obtained results. Finally, Section~\ref{sec:conc} brings our conclusions and future perspectives.

\section{Theoretical background}
\label{sec:theory}

\subsection{Blind source separation}
\label{sec:bse}

\begin{figure}[ht]
\centering
\includegraphics[height=3.5cm]{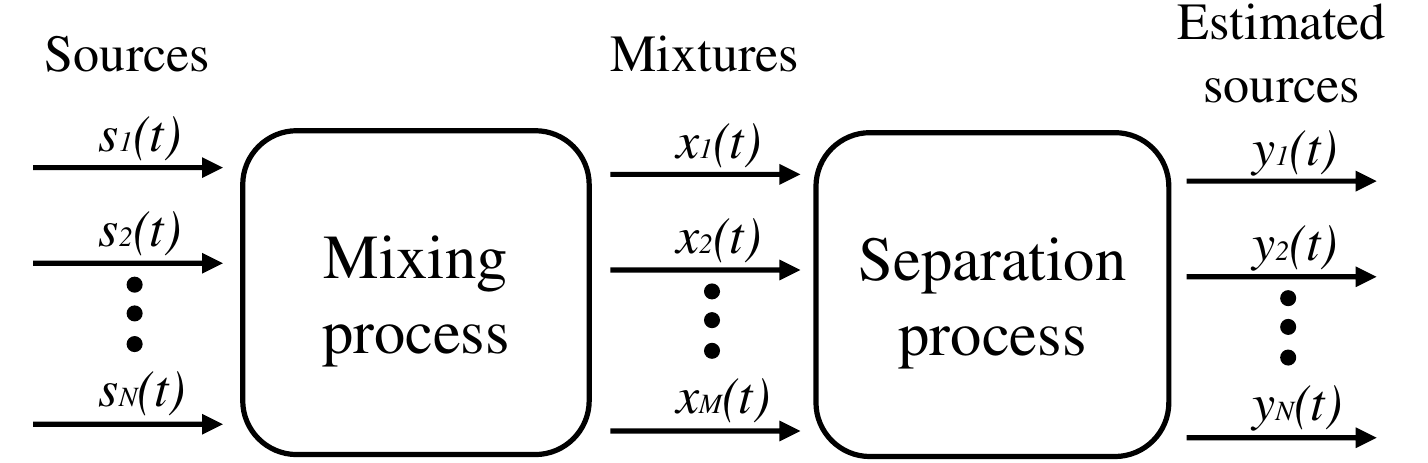}
\caption{BSS problem.}
\label{fig:esqbss}
\end{figure}

Consider a BSS problem as illustrated in Figure~\ref{fig:esqbss}. The aim is to recover a set of (unknown) sources signals $\mathbf{s}(t)=[s_1(t),s_2(t),\ldots,s_N(t)]^T$ based on the observation of $\mathbf{x}(t)=[x_1(t),x_2(t),\ldots,x_M(t)]^T$, which is, in the instantaneous linear case, given by
\begin{equation}
\label{eq:mixture}
\mathbf{x}(t)=\mathbf{A}\mathbf{s}(t) + \mathbf{r}(t),
\end{equation}
where $\mathbf{A} \in \mathbb{R}^{M \times N}$ is the (unknown) mixing matrix and $\mathbf{r}(t)$ represents an additive white Gaussian noise (AWGN) term\footnote{It is worth noting that we consider in this paper the determined case, i.e., $M=N$.}. In this context, the separation may be achieved by adjusting a separation matrix $\mathbf{W} \in \mathbb{R}^{N \times M}$ that provides the set of retrieved signals 
\begin{equation}
\label{eq:separ}
\mathbf{y}(t)=\mathbf{W}\mathbf{x}(t),
\end{equation}
which should be close as possible to $\mathbf{s}(t)$ (up to the well-known permutation and scaling ambiguities~\citep{Comon2010}).

Generally, the procedure to adapt $\mathbf{W}$ involves the solution of a mono-objective optimization problem, given by\footnote{In this work, we consider the minimization of cost functions. However, if $J(\cdot)$ must be maximized, the resulting problem can be transformed into minimization by applying the simple transformation $\max \, J(\cdot)=\min \, -J(\cdot)$.}
\begin{equation}
\label{eq:bssmono}
\begin{array}{ll}
\displaystyle\min_{\mathbf{W}} & \displaystyle J(\mathbf{W}) \\
\end{array}
\end{equation}
whose cost function\footnote{We define $J(\mathbf{W})$ as a function of $\mathbf{W}$ because it is related to a property of the retrieved signals $\mathbf{y}(t)$, which is obtained through~\eqref{eq:separ} given a separation matrix $\mathbf{W}$.} (or separation criterion) $J(\mathbf{W})$ is related to a given property of the sources, such as sparsity, non-Gaussianity or temporality. For example, if one aims at retrieving sparse sources, one may formulate a separation criterion $J_{Spars}(\mathbf{W})$ related to this property and solve the optimization problem expressed in~\eqref{eq:bssmono}, which will lead to the set of estimates $\mathbf{y}(t)$ as sparse as possible.

One may note, therefore, that source separation is strongly dependent on the separation criterion $J(\mathbf{W})$ in representing the desired property. In this context, when $J(\mathbf{W})$ ensures a perfect separation of the source signals, one may say that it acts as a contrast function~\citep{Comon2010}. However, this is not always possible in practical situations, since one may have, for example, interference of random noise.

Another remarkable aspect in solving~\eqref{eq:bssmono} is that one achieves a single solution for the BSS problem, i.e. a single estimate of the set of sources. However, in real applications, such as biomedical signal processing, it may be desirable to have more than one possible estimate in order to make a decision. Regarding this situation, this paper proposes, differently from~\eqref{eq:bssmono}, an approach based on multi-objective optimization, which leads to a set of optimal solutions for the problem. The next section introduces the basic aspects of multi-objective optimization.

\subsection{Multi-objective optimization}
\label{sec:moo}

In mono-objective optimization problems, the model is given by
\begin{equation}
\label{eq:mono}
\begin{array}{ll}
\displaystyle\min_{\mathbf{p}} & \displaystyle f(\mathbf{p}) \\
\text{s.t.} & \textbf{p} \in \Omega \\
\end{array}
\end{equation}
where $f(\textbf{p})$ is the cost function, $\textbf{p}=(p_1, p_2, \ldots , p_L)^T$ is the vector of $L$ variables and $\Omega$, a subset of $\mathbb{R}^L$, is the feasible region\footnote{We refer to feasible region as the region in which the variables satisfy all constraints that may be included in the optimization problem.}. Therefore, for each solution $\textbf{p} \in \mathbb{R}^L$, there is a corresponding $f(\textbf{p}) \in \mathbb{R}$.

However, when there are more than one cost function to be optimized, we may formulate a multi-objective optimization problem~\citep{Miettinen1999}. Let $\mathbf{f}(\mathbf{p}) = \left[f_1(\mathbf{p}), f_2(\mathbf{p}), \ldots, f_K(\mathbf{p})\right]$ be the vector that represents the set of $K$ cost functions. The resulting model can be expressed as follows:
\begin{equation}
\label{eq:moo}
\begin{array}{ll}
\displaystyle\min_{\mathbf{p}} & \displaystyle\left[f_1(\mathbf{p}), f_2(\mathbf{p}), \ldots, f_K(\mathbf{p})\right] \\
\text{s.t.} & \textbf{p} \in \Omega \\
\end{array}
\end{equation}
where $\textbf{p}=(p_1, p_2, \ldots , p_L)^T$ is the vector of $L$ variables and $\Omega$, a subset of $\mathbb{R}^L$, is the feasible region. In this formulation, for each solution $\textbf{p} \in \mathbb{R}^L$, there is an $\mathbf{f}(\textbf{p}) \in \mathbb{R}^K$.

The main difference between mono and multi-objective approaches is that the latter involves the optimization of a vector-valued cost function. Therefore, the notion of optimality changes, since a solution $\textbf{p}'$ for~\eqref{eq:mono} is the one that minimizes a single value $f(\mathbf{p})$ and a solution $\textbf{p}''$ for~\eqref{eq:moo} is the one that minimizes all $f_i(\mathbf{p})$, $i=1, \ldots, K$, simultaneously. However, one often finds conflicts between these cost functions and, consequently, it is impossible to find a single solution that minimizes all of them. In these situations, the solution of~\eqref{eq:moo} leads to a set of non-dominated points~\citep{Deb2001,Miettinen1999}, defined, in a minimization problem, as follows:

\begin{defn}
\label{def:pareto}
A solution $\mathbf{p}$ is a non-dominated solution if there is no other solution $\mathbf{p}'$ such that $f_i(\mathbf{p}') \leq f_i(\mathbf{p})$ for all $i=1, 2, \ldots, K$ and $f_j(\mathbf{p}') < f_j(\mathbf{p})$ for at least one index $j$.
\end{defn}

\begin{figure}[ht]
\centering
\includegraphics[height=6.0cm]{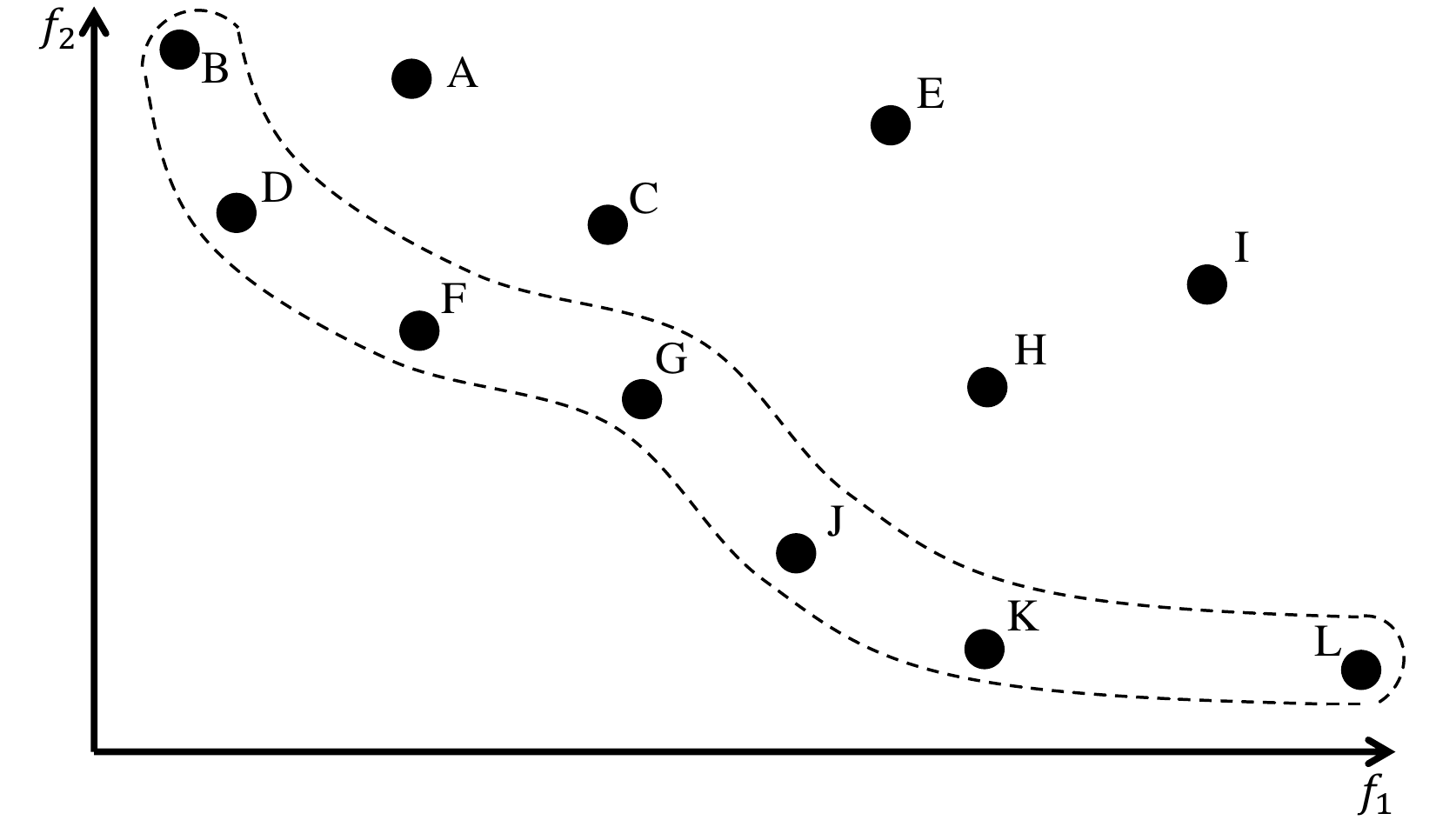}
\caption{Objective space and the non-dominated set in a minimization problem.}
\label{fig:pareto}
\end{figure}

In order to illustrate the non-dominated solutions, consider the objective space illustrated in Figure~\ref{fig:pareto}. Aiming at minimizing both cost functions, one can find the non-dominated solutions represented by the ones highlighted by the dashed line. Comparing these solutions, one may note trade-offs between them, i.e. if a solution is better than another with respect to a specific criterion, it will be necessarily worse with respect to another criterion. Furthermore, among the solutions in the non-dominated set, there are those that minimizes each criterion individually (solutions B and L, in this case). 

\section{Multi-objective BSS approach}
\label{sec:mooapr}

\begin{figure}[ht]
\centering
\includegraphics[height=7.5cm]{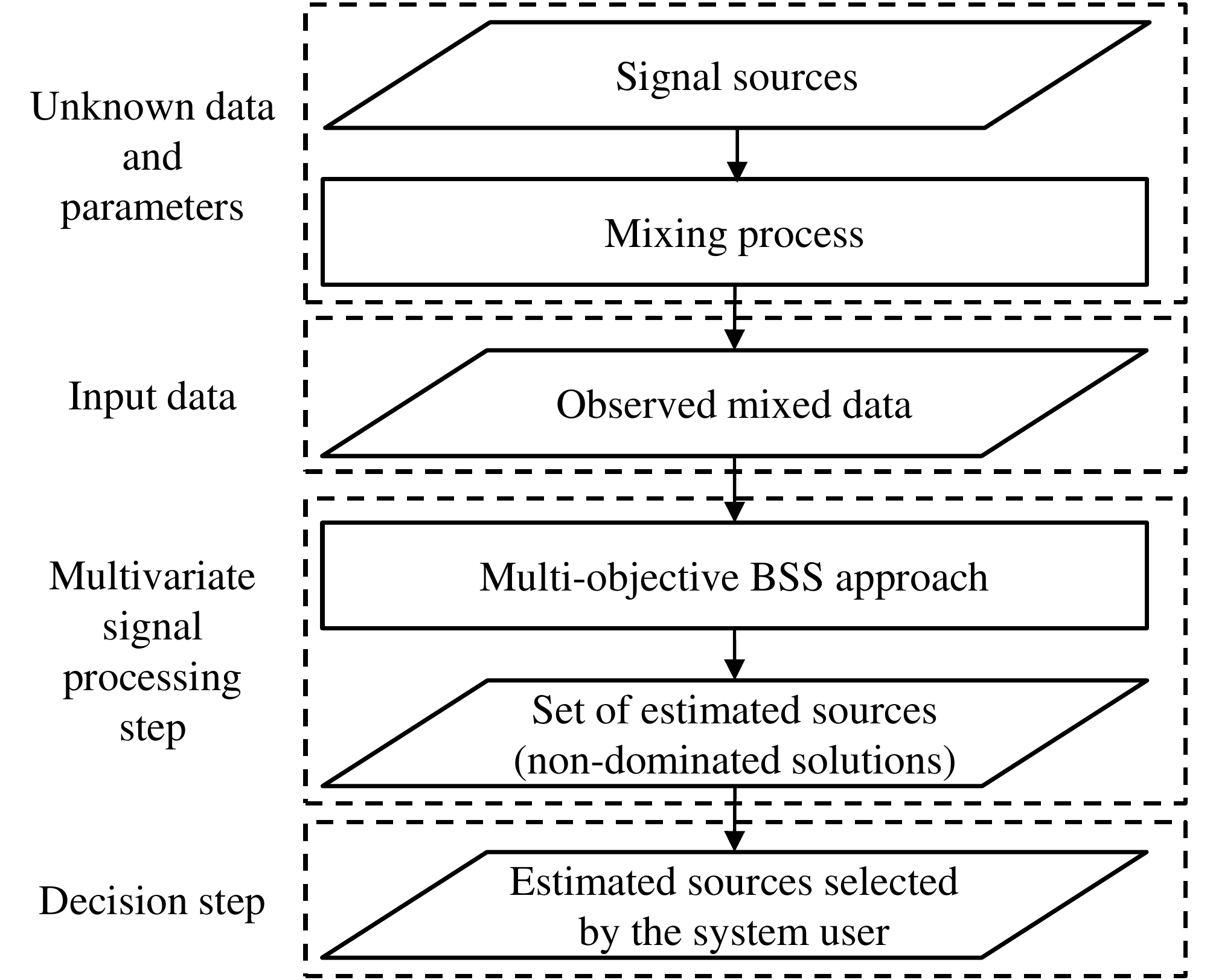}
\caption{Multi-objective blind source separation system.}
\label{fig:scheme}
\end{figure}

In this section, we present the proposed Multi-Objective Blind Source Separation (MO-BSS) approach used to deal with BSS problems. A general scheme is presented in Figure~\ref{fig:scheme}. Based on the observed mixed data obtained by the unknown source signals and mixing process, one firstly applies the MO-BSS approach in order to obtain the set of non-dominated solutions. Given this set, which represents a set of estimated sources, the last step comprises the selection of a proper one according to a subjective knowledge on the problem.

\subsection{Main reasons for the application of the multi-objective optimization models to deal with BSS problems}
\label{subsec:model}

We are interested in BSS problems for which there is no separation criterion that guarantees a perfect separation of the source signals, which is the case found in practical situations. However, we assume that we have more than one prior information about the sources. Therefore, we can build more than one criterion related to their properties in order to deal with the separation problem.

As mentioned in Section~\ref{sec:intro}, some works take into account more than one property of the sources and combine them into a single optimization criterion. A possible formulation is to consider weighting factors associated with each property. The resulting model is given by
\begin{equation}
\label{eq:mono_w}
\begin{array}{ll}
\displaystyle\min_{\mathbf{W}} & \displaystyle \lambda_1 J_1(\mathbf{W}) + \lambda_2 J_2(\mathbf{W}) + \ldots + \lambda_K J_K(\mathbf{W}) \\
\end{array}
\end{equation}
where $\boldsymbol{\lambda}=\left(\lambda_1, \lambda_2, \ldots, \lambda_K\right)$ is the weight vector. By solving~\eqref{eq:mono_w} for a predetermined $\boldsymbol{\lambda}$, one obtains an optimal separation matrix $\hat{\mathbf{W}}$ which leads to a single estimation of the sources. A remarkable aspect in this formulation is that, if one varies the weights $\boldsymbol{\lambda}$, one may achieve different solutions for~\eqref{eq:mono_w}. As a consequence, one may achieve different estimations for the source signals.

\begin{figure}[ht]
\centering
\subfloat[Sources.]{\includegraphics[width=3.0in]{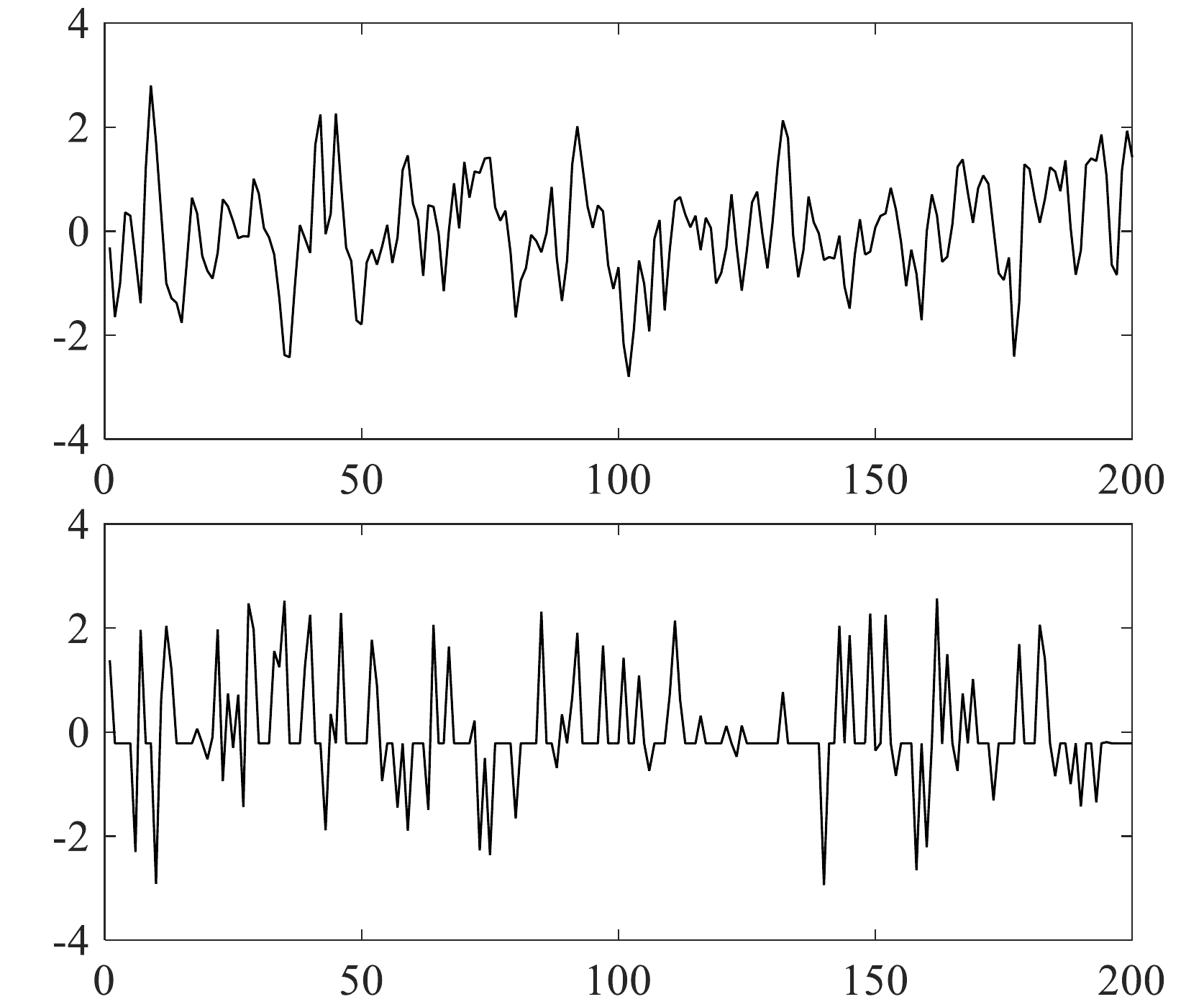}
\label{fig:lambda_s}}
\hfil
\subfloat[Mixtures.]{\includegraphics[width=3.0in]{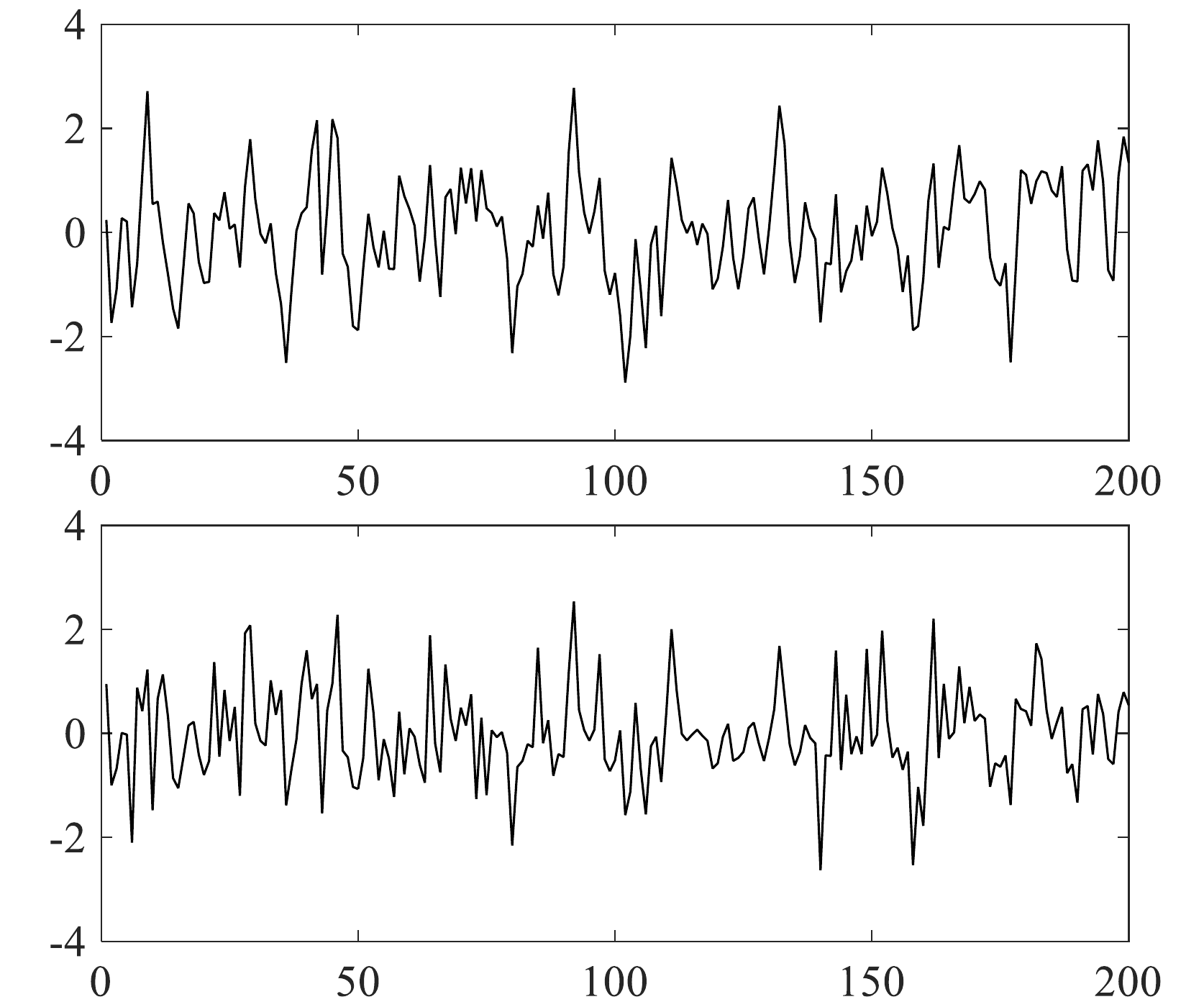}
\label{fig:lambda_m}}
\caption{Example of sources and mixtures.}
\label{fig:lambda_data}
\end{figure}

\begin{figure}[ht]
\centering
\subfloat[$\lambda_1=0.3$ and $\lambda_2 = 0.7$.]{\includegraphics[width=3.0in]{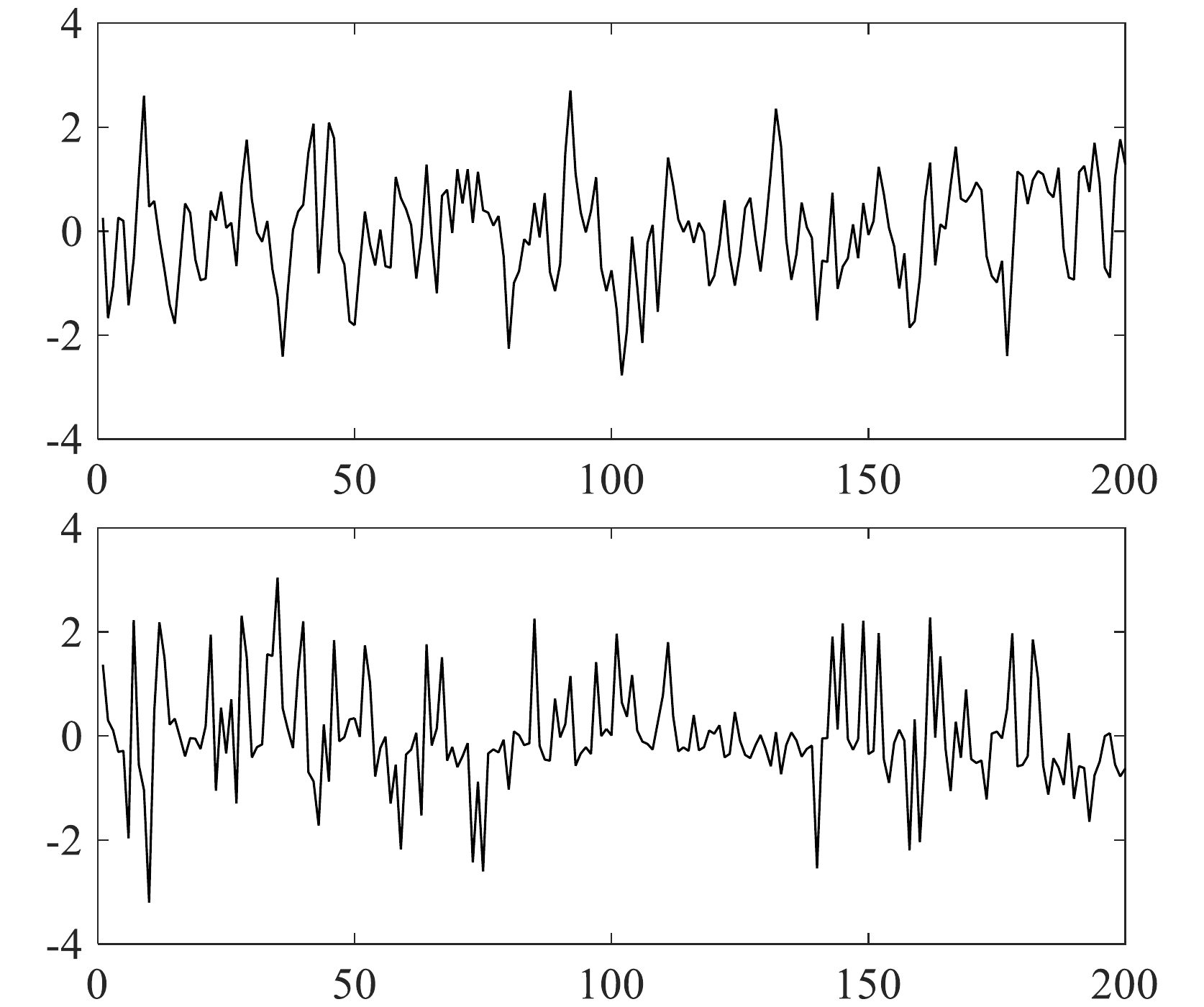}
\label{fig:lambda_03}}
\hfil
\subfloat[$\lambda_1=0.7$ and $\lambda_2 = 0.3$.]{\includegraphics[width=3.0in]{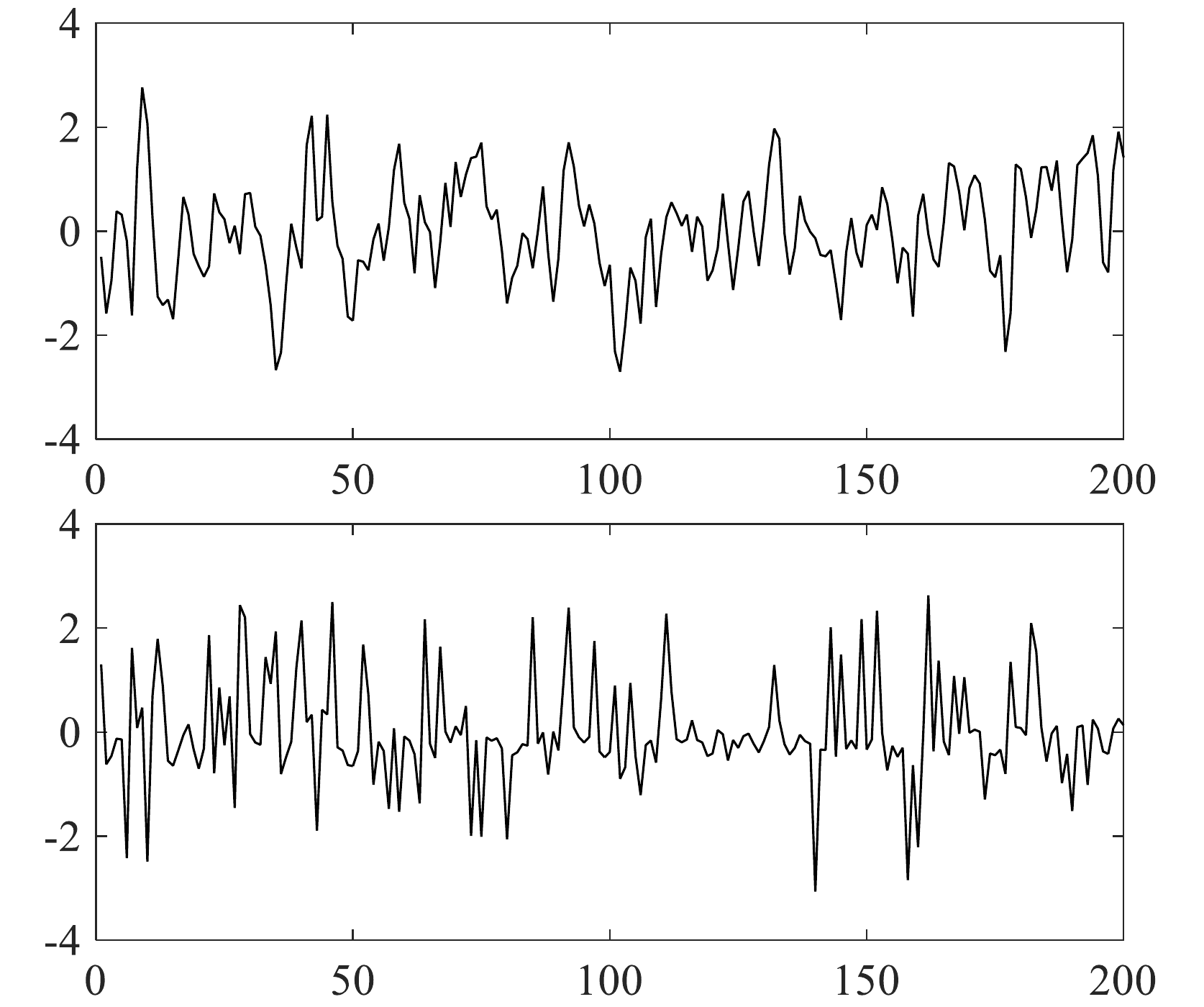}
\label{fig:lambda_07}}
\caption{Estimated sources for different weights.}
\label{fig:lambda_est}
\end{figure}

In order to illustrate the aforementioned situation, consider the set of (two) sources and the set of (two) mixtures presented in Figures~\ref{fig:lambda_s} and~\ref{fig:lambda_m}, respectively. Suppose that we formulate an optimization model composed by two criteria related to two properties of the sources, for instance, sparsity ($J_{Spars}$) and decorrelation ($J_{Decorr}$). The resulting model is given by
\begin{equation}
\label{eq:ex_lambda}
\begin{array}{ll}
\displaystyle\min_{\mathbf{W}} & \displaystyle \lambda_1 J_{Spars}(\mathbf{W}) + \lambda_2 J_{Decorr}(\mathbf{W}) \\
\end{array}
\end{equation}
By solving~\eqref{eq:ex_lambda} for $\left[\lambda_1, \lambda_2 \right] = \left[0.3, 0.7 \right]$ and for $\left[\lambda_1, \lambda_2 \right] = \left[0.7, 0.3 \right]$, we find the estimated sources illustrated in Figures~\ref{fig:lambda_03} and~\ref{fig:lambda_07}, respectively. One may note that both set of estimates are different for each weight vector considered in~\eqref{eq:ex_lambda}, which may be an inconvenient, since one does not know in advance what is the weight vector that we need to consider in order to achieve a proper estimation of the source signals.

Aiming at taking into account a set of informations about the source signals but avoiding an \textit{a priori} definition of the set of weights $\boldsymbol{\lambda}$, this paper proposes the application of a multi-objective optimization approach to deal with BSS problems (MO-BSS approach). Let $\mathbf{J}(\mathbf{W}) = \left[J_1(\mathbf{W}), J_2(\mathbf{W}), \ldots, J_K(\mathbf{W})\right]$ be the vector that represents the set of $K$ separation criteria associated with different properties of the sources. Differently from~\eqref{eq:mono_w}, the resulting model is given by
\begin{equation}
\label{eq:moo_bss}
\begin{array}{ll}
\displaystyle\min_{\mathbf{W}} & \displaystyle \left[J_1(\mathbf{W}), J_2(\mathbf{W}), \ldots, J_K(\mathbf{W})\right] \\
\end{array}
\end{equation}
In this formulation, the solution of~\eqref{eq:moo_bss} leads to a set of non-dominated estimates which can be used as a basis for the system user to select the best one according to his/her subjective knowledge on the problem.

\begin{figure}[ht]
\centering
\includegraphics[height=7.0cm]{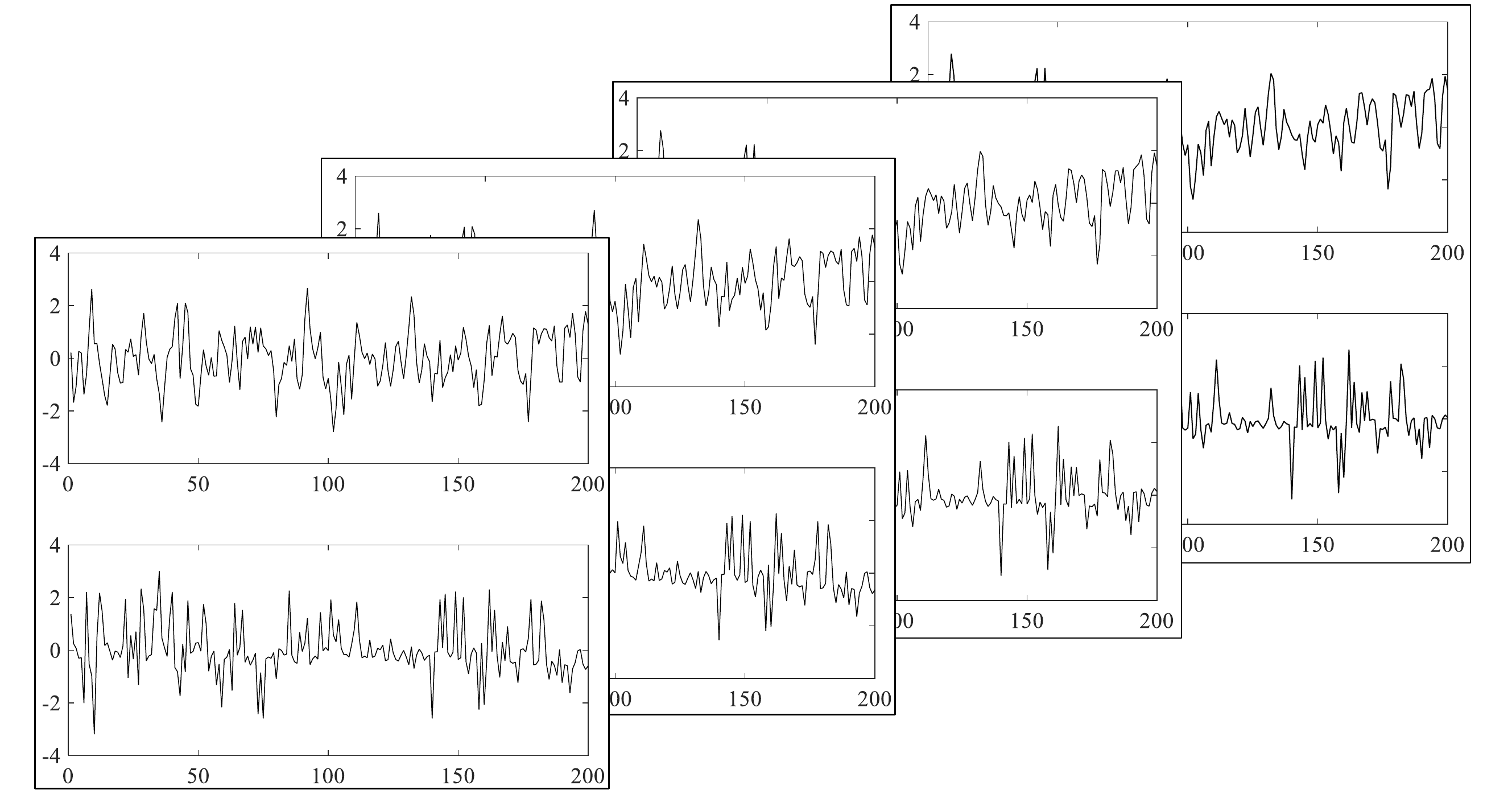}
\caption{Set of optimal solutions.}
\label{fig:lambda_moo}
\end{figure}

If one considers the aforementioned example, the solution of the multi-objective problem
\begin{equation}
\label{eq:ex_lambda_moo}
\begin{array}{ll}
\displaystyle\min_{\mathbf{W}} & \displaystyle \left[J_{Spars}(\mathbf{W}), J_{Decorr}(\mathbf{W})\right] \\
\end{array}
\end{equation}
leads to the set of estimates illustrated in Figure~\ref{fig:lambda_moo}. Therefore, compared to~\eqref{eq:mono_w}, the multi-objective formulation allows us an \textit{a posteriori} selection of a set of estimates according to an additional information or subjective knowledge on the problem.

\subsection{Multi-objective optimization technique}
\label{subsec:tech}

In the literature, one may find several techniques to estimate the non-dominated set. Commonly, they are divided into two main groups: the classical methods, such as the weighted sum and the $\epsilon$-constraint~\citep{Miettinen1999}, and the techniques founded on metaheuristics, such as those based on evolutionary algorithms~\citep{Deb2001}. With respect to the classical methods, generally, they require some mathematical analysis on the cost function and/or constraints. For example, the weighted sum method, which comprises the determination of the non-dominated set through successive solutions of~\eqref{eq:mono_w} (for different values of $\boldsymbol{\lambda}$), ensures that all possible non-dominated solution can be obtained only if the objective space (space of cost functions) is convex~\citep{Miettinen1999}. On the other hand, evolutionary algorithms may be applied to determine the non-dominated set without assuming, for example, convexity of the objective space. Therefore, these algorithms may be used in a wide variety of applications~\citep{Deb2008}. A price to be paid is that they do not guarantee optimality, since they are based on heuristics. 

Different problems in BSS can be exploited by means of different separation criteria. Therefore, several optimization problem may be formulated. In this context, if one decides to use classical methods, in each optimization model, one should perform a mathematical analysis to guarantee that the considered method can provide the desired results. However, since in this paper we propose a general expert system that can be used in any BSS problem (under the adjustment of the cost functions, of course), we consider the use of evolutionary algorithms. Specifically, we apply an improved version of the well-known Strenght Pareto Evolutionary Algorithm~\citep{Zitzler1998}, called SPEA2.

Introduced by Zitzler, Laumanns and Thiele~\citeyearpar{Zitzler2001}, the steps of SPEA2 are described in Algorithm~\ref{alg:spea}. Before starting, one randomly generates an initial population $\mathbf{Z}$, which comprises, in the BSS problem, a set of possible candidates (or individuals) for the separation matrix $\mathbf{W}$. In the \textit{fitness assignment} step, for each candidate in $\mathbf{Z}$ and in the external set $\overline{\mathbf{Z}}$, which is empty in the first iteration but will store the best candidates achieved so far, one finds the corresponding estimated sources (according to~\eqref{eq:separ}), calculates the criteria values $J_1(\mathbf{W}), J_2(\mathbf{W}), \ldots, J_K(\mathbf{W})$ and defines a quality measure (known as \textit{fitness}) based on the dominance relation among the individuals. In the second step, one updates the external set $\overline{\mathbf{Z}}$ with the best $\overline{B}$ candidates found so far, where $\overline{B}$ is the predefined external set size. If the maximum number of iterations $G$ is achieved (or other stopping criteria is satisfied), the algorithm terminates and the set of estimates are obtained by considering the separating matrices stored in $\overline{\mathbf{Z}}$. Otherwise, one moves to the forth step and determines the $B$ (predefined population size) candidates that will be submitted to the next step by means of a binary tournament selection with replacement performed in $\overline{\mathbf{Z}}$. Finally, in the \textit{variation} step, one applies evolutionary operators (crossover and mutation) on the $B$ selected candidates in order to generate the new population $\mathbf{Z}$ ($\alpha$\% by crossover and $(1-\alpha)$\% by mutation), which will be submitted to the first step to restart the algorithm.

An important aspect in the SPEA2 algorithm is the diversity of the non-dominated set. Besides storing the best candidates after each iteration, it is also required to update the external set with the ones that are more separated from each other. This promotes diversity and preserves the candidates in the limits of the non-dominated set (optimal solutions for each criterion taken individually). We refer the reader to~\citep{Zitzler2001} for further details about each step of SPEA2 algorithm.

\begin{algorithm}
    \caption{SPEA2}
    \label{alg:spea}
		\begin{algorithmic}
				\STATE \textbf{Input:} Initial population $\mathbf{Z}$, empty external set $\overline{\mathbf{Z}}$, population size $B$, external set size $\overline{B}$, maximum number of iterations $G$ and crossover rate $\alpha$.
				\STATE \textbf{Output:} External set $\overline{\mathbf{Z}}$.
				\STATE Set $g=1$.
				\WHILE{$g \leq G$}
						\STATE \textit{Step 1: Fitness assignment}. For each candidate in $\mathbf{Z}$ and $\overline{\mathbf{Z}}$, calculate the fitness measure.
						\STATE \textit{Step 2: Selection}. Based on the fitness measures, update $\overline{\mathbf{Z}}$ with the best $\overline{B}$ candidates.
						\IF{$g = G$}
								\STATE \textit{Step 3: Termination}. Define the external set $\overline{\mathbf{Z}}$ as the non-dominated set and stop the algorithm.
						\ENDIF
						\STATE \textit{Step 4: Mating selection}. Perform a binary tournament selection in $\overline{\mathbf{Z}}$ and select $B$ candidates for the next step.
						\STATE \textit{Step 5: Variation}. Apply crossover and mutation on the selected candidates and generate the new population $\mathbf{Z}$. Set $g = g + 1$.
				\ENDWHILE
    \end{algorithmic}
\end{algorithm}

\subsection{On the selection of proper estimated sources}
\label{subsec:selec}

After running the SPEA2 algorithm, one obtains the non-dominated set, which leads to a set of possible estimates for the BSS problem. Since the problem is blind, it is not possible to determine the best one in this set. Therefore, an important issue in our proposed approach is how to select a proper source estimate among all estimates in the non-dominated set.

An approach that we propose in this paper and that will be used in our experiments comprises two steps. In the first step, we order the non-dominated solutions according to their evaluations in a separation criterion. For instance, if we consider the multi-objective optimization problem illustrated in Figure~\ref{fig:pareto} and the separation criterion expressed by $f_1$, we obtain the following ordering: B, D, F, G, J, K and L. Based on this ordering, the second step comprises the visualization of the (ordered) non-dominated solutions and the selection of a proper one according to an additional information or a subjective knowledge on the problem. If we consider the results illustrated in Figure~\ref{fig:pareto}, one may note that estimates B and L are more associated with the properties expressed by $f_1$ and $f_2$, respectively. Varying the solutions from B to L, one may select a proper one based on another desirable characteristic of the sources (not necessarily a property that is possible to express in a mathematical formula).

\section{Numerical experiments}
\label{sec:exp}

\subsection{Dataset}
\label{subsec:expbss}

\begin{figure}[ht]
\centering
\subfloat[Sources.]{\includegraphics[width=2.8in]{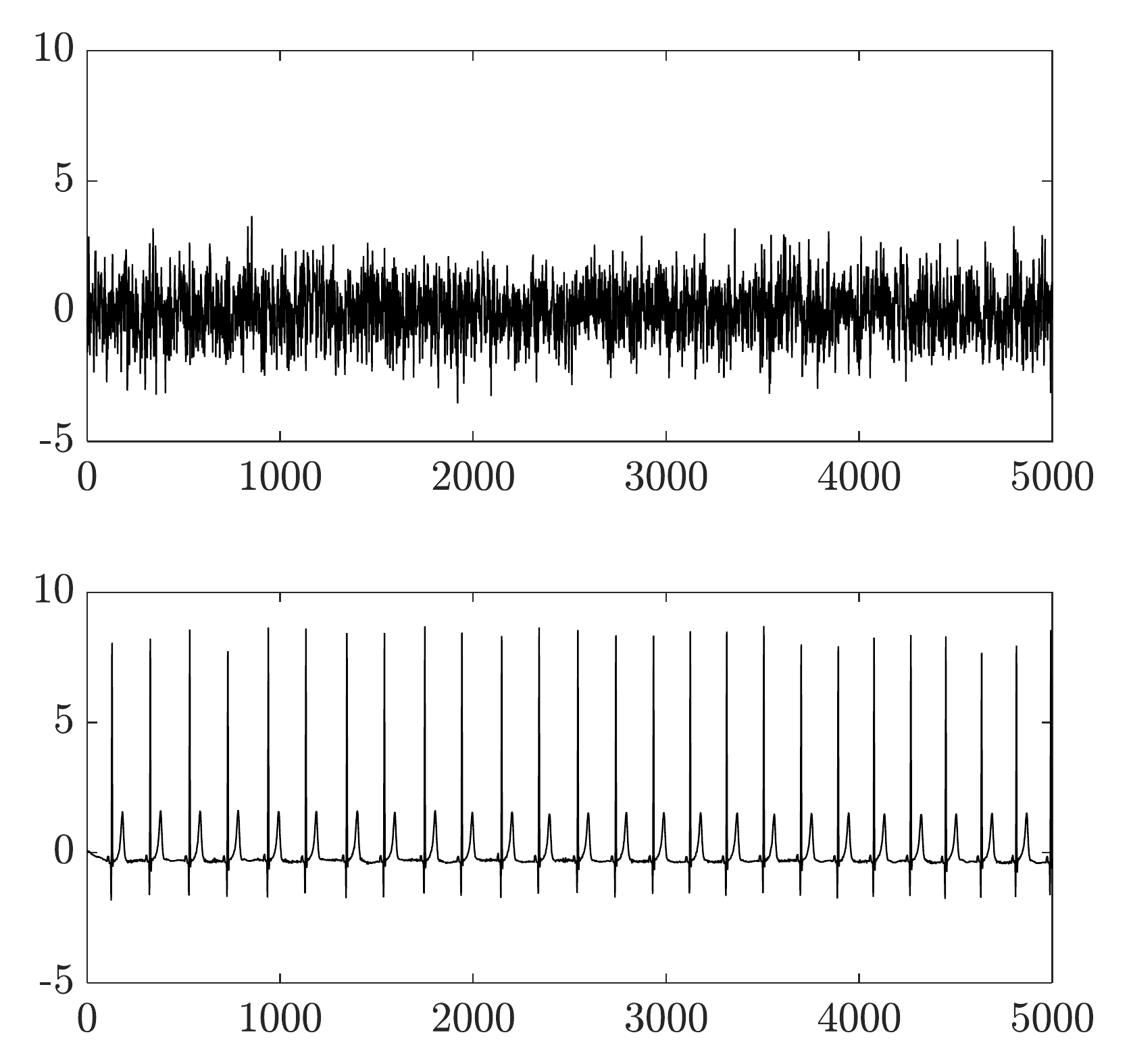}
\label{fig:sourc_bss}}
\hfil
\subfloat[Observed mixtures.]{\includegraphics[width=2.8in]{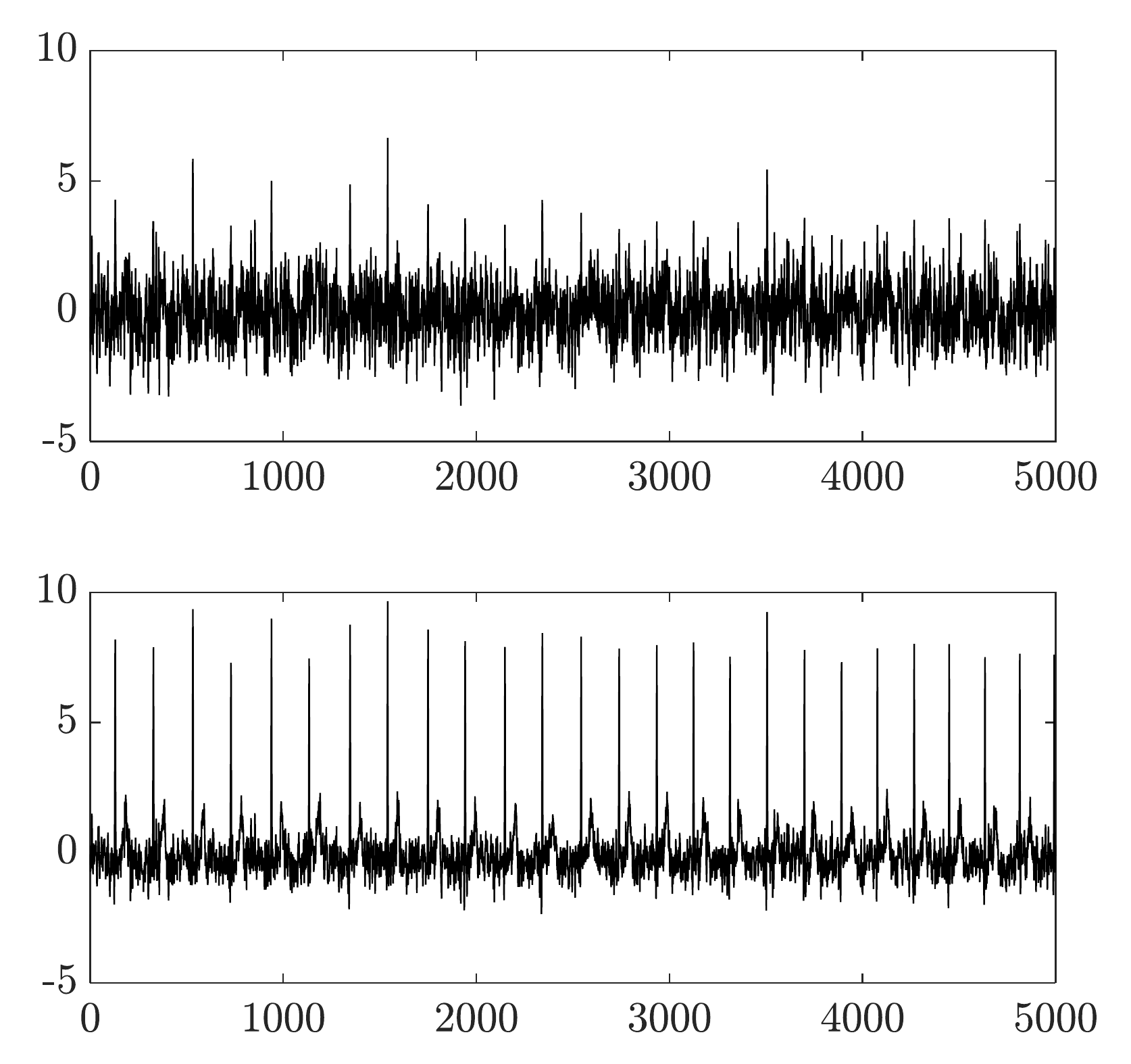}
\label{fig:mix_bss}}
\caption{Source signals and observed mixtures.}
\label{fig:sourc_mix_bss}
\end{figure}

In order to illustrate the application of the proposed MO-BSS approach, we performed numerical experiments in a BSS problem comprised by two sources, illustrated in Figure~\ref{fig:sourc_bss}, which were obtained from the ABio7 dataset in the ICALAB toolbox~\citep{Cichocki}. One may note that one source is a ECG signal. After applying a synthetic linear mixture (without noise) according to the mixing matrix
$$\mathbf{A}=\left[ \begin{array}{cc}
1 & 0.5 \\ 
0.5 & 1\\
\end{array} \right],$$
we obtained the observed mixtures presented in Figure~\ref{fig:mix_bss}.

\subsection{Adopted separation criteria}
\label{subsubsec:bsscrit}

In this experiment, we aim at separating the two sources. Assuming that we have the information that one source is an ECG signal, we can also use both time-correlation and sparsity characteristics in order to formulate the optimization criteria (even that only one has both properties).

\begin{figure}[ht]
\centering
\includegraphics[height=5.0cm]{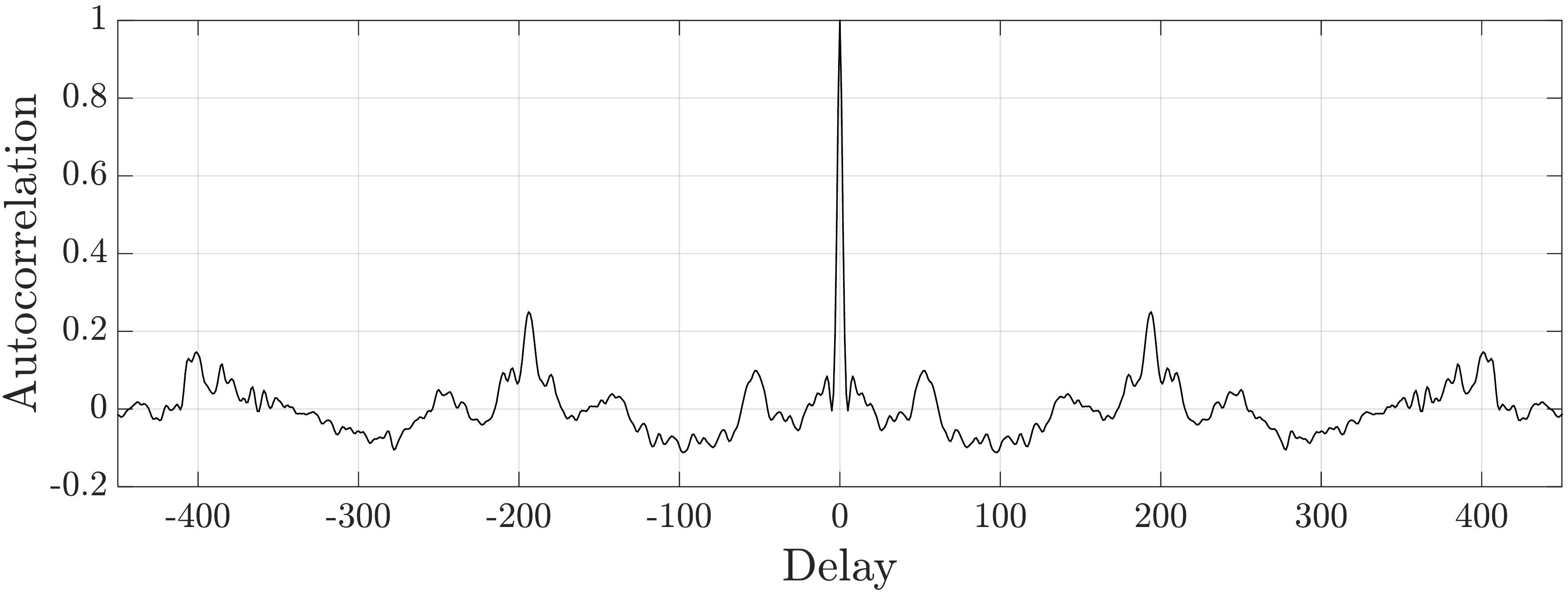}
\caption{Autocorrelation function of the second mixture.}
\label{fig:bssautoc}
\end{figure}

By taking into account the time-correlation characteristic, one may consider the maximization of the autocorrelation coefficient of both estimated sources. Therefore, the optimization criterion adapted to a minimization problem is given by
\begin{equation}
J_1(\mathbf{W})=-\sum_{i=1}^2 \left|E\left[\mathbf{w}_i^T\mathbf{x}(t)\mathbf{x}(t-\tau)^T\mathbf{w}_i\right]\right|,
\end{equation}
where $\mathbf{w}_i^T$ represents the row $i$ of the separation matrix $\mathbf{W}$ associated with the estimation of the source $i$. One may note that the calculation of $J_1(\mathbf{W})$ depends on a predefined delay $\tau$. This parameter may be found directly from the autocorrelation function of the mixture in which the temporal structure is, visually, more evident. In this case, by applying the autocorrelation function on the second mixture, illustrated in Figure~\ref{fig:bssautoc}, we verify that the maximum is achieved (for $\tau \neq 0$) when $\tau=193$.

Based on the second information, we build a separation criterion that takes into account the sparsity of the retrieved sources through the $\ell_1$-norm minimization~\citep{Nadalin2010}. In order to avoid scaling problems, we may normalize the the $\ell_1$-norm, which leads to the following optimization criterion:
\begin{equation}
J_2(\mathbf{W})=\sum_{i=1}^2 \frac{\left\|\mathbf{w}_i^T \mathbf{x}(t)\right\|_1}{\left\|\mathbf{w}_i^T \mathbf{x}(t)\right\|_2}.
\end{equation}

It is worth noting that both considered criteria have different characteristics. Sparsity is related to vectors for which most of elements are close to zero. On the other hand, the autocorrelation exploits the temporal structure of the retrieved sources. Therefore, solving the multi-objective problem
\begin{equation}
\label{eq:moo_exp_bss}
\begin{array}{ll}
\displaystyle\min_{\mathbf{W}} & \displaystyle \left[- \sum_{i=1}^2 \left|E\left[\mathbf{w}_i^T\mathbf{x}(t)\mathbf{x}(t-\tau)^T\mathbf{w}_i\right]\right| \, , \,  \sum_{i=1}^2 \frac{\left\|\mathbf{w}_i^T \mathbf{x}(t)\right\|_1}{\left\|\mathbf{w}_i^T \mathbf{x}(t)\right\|_2} \right], \\
\end{array}
\end{equation}
we expect to obtain non-dominated solutions representing the trade-off between both properties.

\subsection{SPEA2 parameters and additional adjustments}
\label{subsec:spea2par}

Since the addressed BSS problem is composed by two source signals, one aims at retrieving a separation matrix $\mathbf{W} \in \mathbb{R}^{2 \times 2}$. Therefore, each individual in SPEA2 algorithm is composed by a vector of four variables. With respect to the parameters input, they were defined based on experimental tests which, as will be discussed later on this paper, lead to acceptable convergence and diversity of the obtained non-dominated set. For instance, we considered the following: population size $\mathbf{Z}=100$, external set size $\tilde{\mathbf{Z}}=50$, maximum number of iterations $G=60$ and crossover rate $\alpha=50$.

It is worth mentioning that, since the ECG signal presents both properties expressed in the adopted criteria, the SPEA2 described in Algorithm~\ref{alg:spea} could lead to separation matrices $\mathbf{W}$ that provide estimates for the same source (i.e. the ECG signal). Therefore, one should adjust the algorithm by including a feasibility constraint. This can be achieved by adding a penalization in the infeasible individuals when calculating the fitness measure. For instance, if an individual leads to estimates whose Pearson coefficient is greater than 0.2 (a low level of correlation between the signals), one adds $10^5$ in both criteria. With this penalization, the quality of this individual, expressed by the fitness measure, will decrease. As a consequence, it will not be selected for the external set.

\subsection{Experiment with linear mixture without noise}
\label{subsec:explin}

In this analysis, we considered a linear mixture, as presented in~\eqref{eq:mixture}, without the additive noise $\mathbf{r}(t)$. In order to verify the quality of the obtained results (the set of retrieved signals), we consider the signal-to-interference ratio (SIR), given by
\begin{equation}
\label{eq:sir}
SIR_i=10 \log \left(\frac{E[s_i(t)^2]}{E[(s_i(t)-y_i(t))^2]}\right),
\end{equation}
and, as a benchmark, the MSE solution\footnote{Since the MSE solution is the one that maximizes the SIR between the source signals and estimated ones, we used it as benchmark for the set of retrieved sources.}, given by:
\begin{equation}
\label{eq:mse_sol}
\begin{array}{ll}
\displaystyle\min_{\mathbf{W}} & \displaystyle E\left[\sum_{i=1}^2{(s_i(t)-y_i(t))^2}\right]. \\
\end{array}
\end{equation}
It is worth noting that the scales of both $s_i(t)$ and $y_i(t)$ must be the same. Therefore, it is necessary to normalize both signals before calculating the SIR.

\begin{figure}[ht]
\centering
\includegraphics[height=6.5cm]{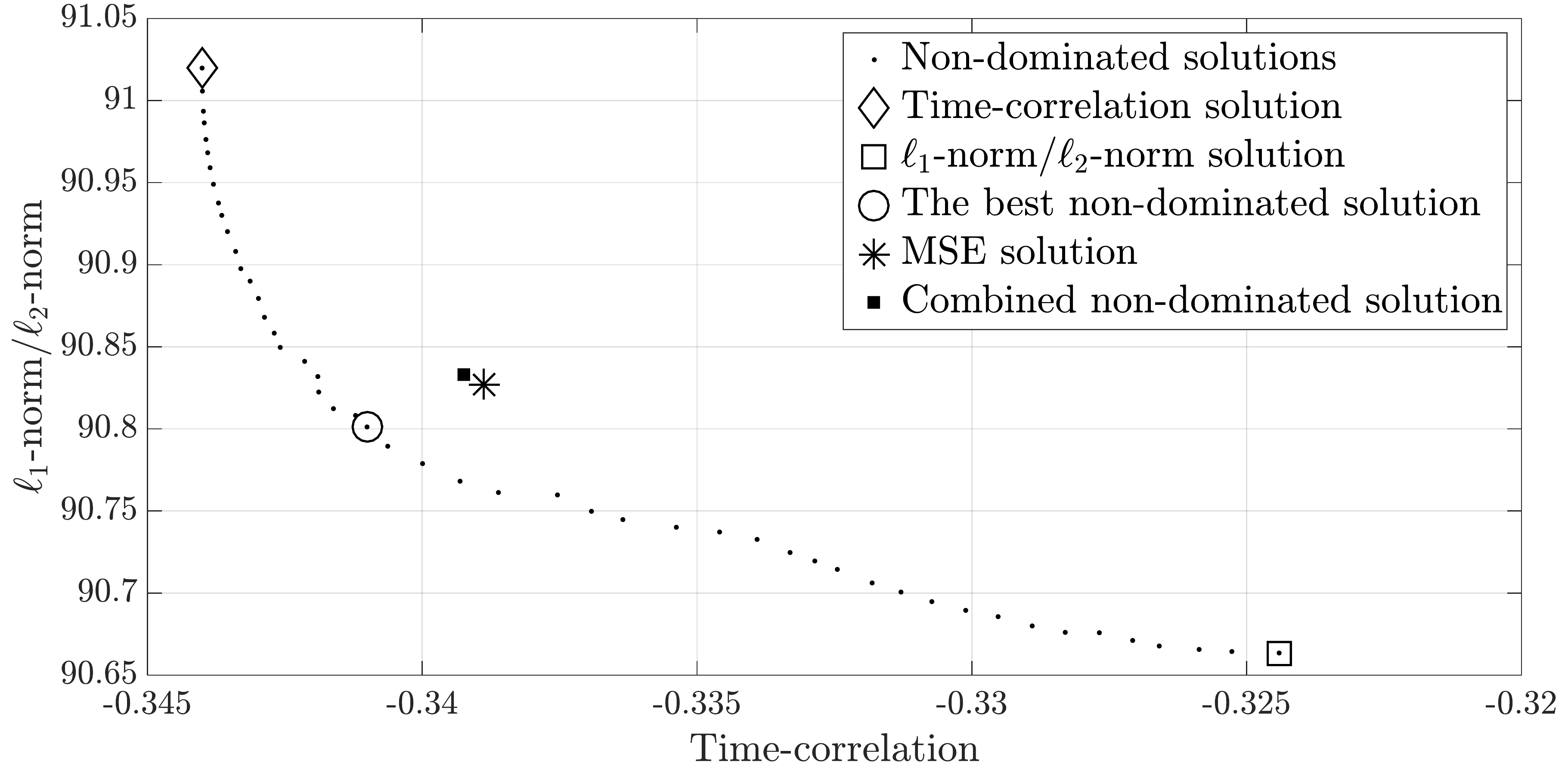}
\caption{Non-dominated set obtained by SPEA2 algorithm and MSE solution.}
\label{fig:bsspareto}
\end{figure}

The non-dominated set obtained by applying the SPEA2 algorithm is shown in Figure~\ref{fig:bsspareto}, as well as the solutions that minimize each criterion individually, the best one in the non-dominated set (which leads to the higher averaged SIR value) and the supervised solution obtained by the MSE approach. The results indicate that a solution in the non-dominated set is closer (in the objective space) to the MSE solution compared to the ones obtained optimizing each criterion individually.

\begin{figure}[ht!]
\centering
\includegraphics[height=7.0cm]{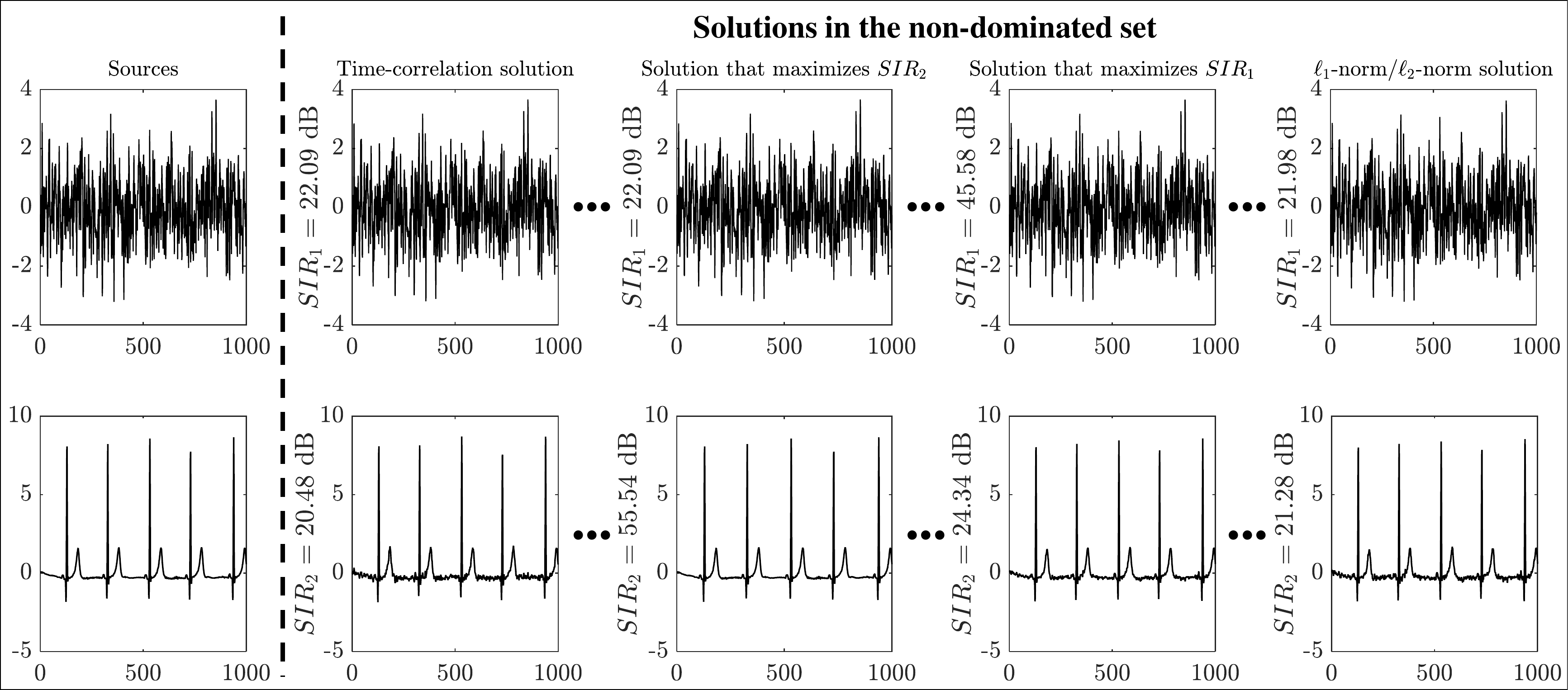}
\caption{Sources and estimations.}
\label{fig:bssest}
\end{figure}

\begin{figure}[ht!]
\centering
\includegraphics[height=7.0cm]{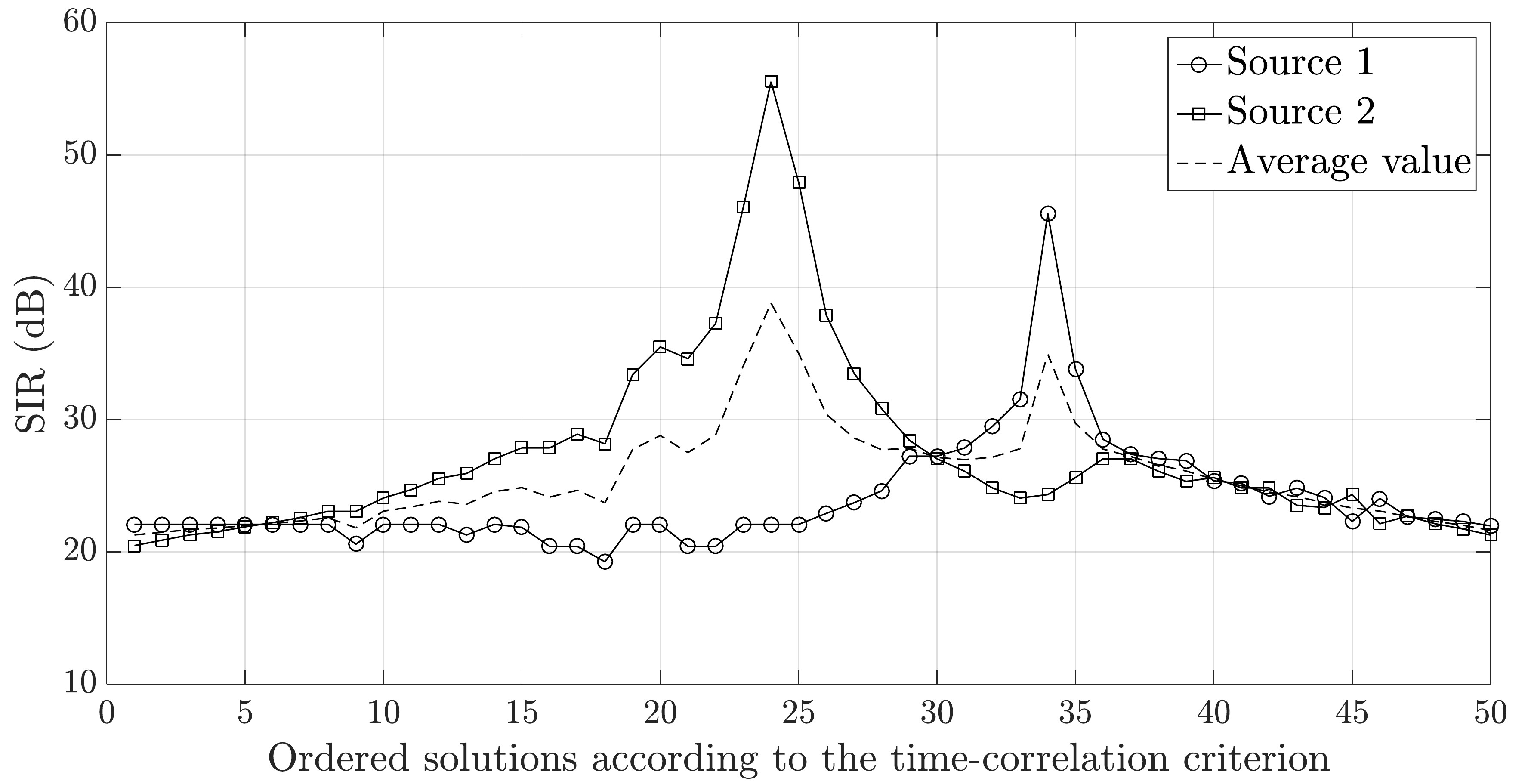}
\caption{SIR values of the non-dominated solutions ordered according to the time-correlation criterion.}
\label{fig:evolsir}
\end{figure}

Figure~\ref{fig:bssest} presents the sources and the retrieved signals for the mono-objective formulations and the solutions in the non-dominated set that maximizes the SIR value of each estimate, as well as the associated SIR values. One remark that the solution that leads to the highest averaged SIR value (the best one in the non-dominated set) corresponds to the solution that maximizes the SIR value for the second source. However, it does not maximizes the SIR value for the first source. Figure~\ref{fig:evolsir} illustrates the SIR values of the non-dominated solutions ordered according to the time-correlation criterion. A remarkable aspect is that, for both sources, the SIR value increases until it reaches a peak and decreases after this point. Moreover, any solution in the non-dominated set is practically better compared to the ones obtained by the mono-objective formulation (the first and the last solutions in Figure~\ref{fig:evolsir}).

Therefore, based on a subjective knowledge on the problem, the system user can select a proper solution among the non-dominated set. Moreover, he/she is also able to combine different solutions in order to achieve a better set of estimates. In the best scenario, he/she may combine the two solutions that maximize the SIR value of the two sources ($SIR_1=45.58$ and $SIR_2=55.54$) and, therefore, achieve estimates (combined non-dominated solution) that are even closer to the MSE solution, as illustrated in Figure~\ref{fig:bsspareto}.

\subsection{Convergence analysis of the non-dominated solutions}

\begin{figure}[ht!]
\centering
\subfloat[First iterations.]{\includegraphics[width=5.5in]{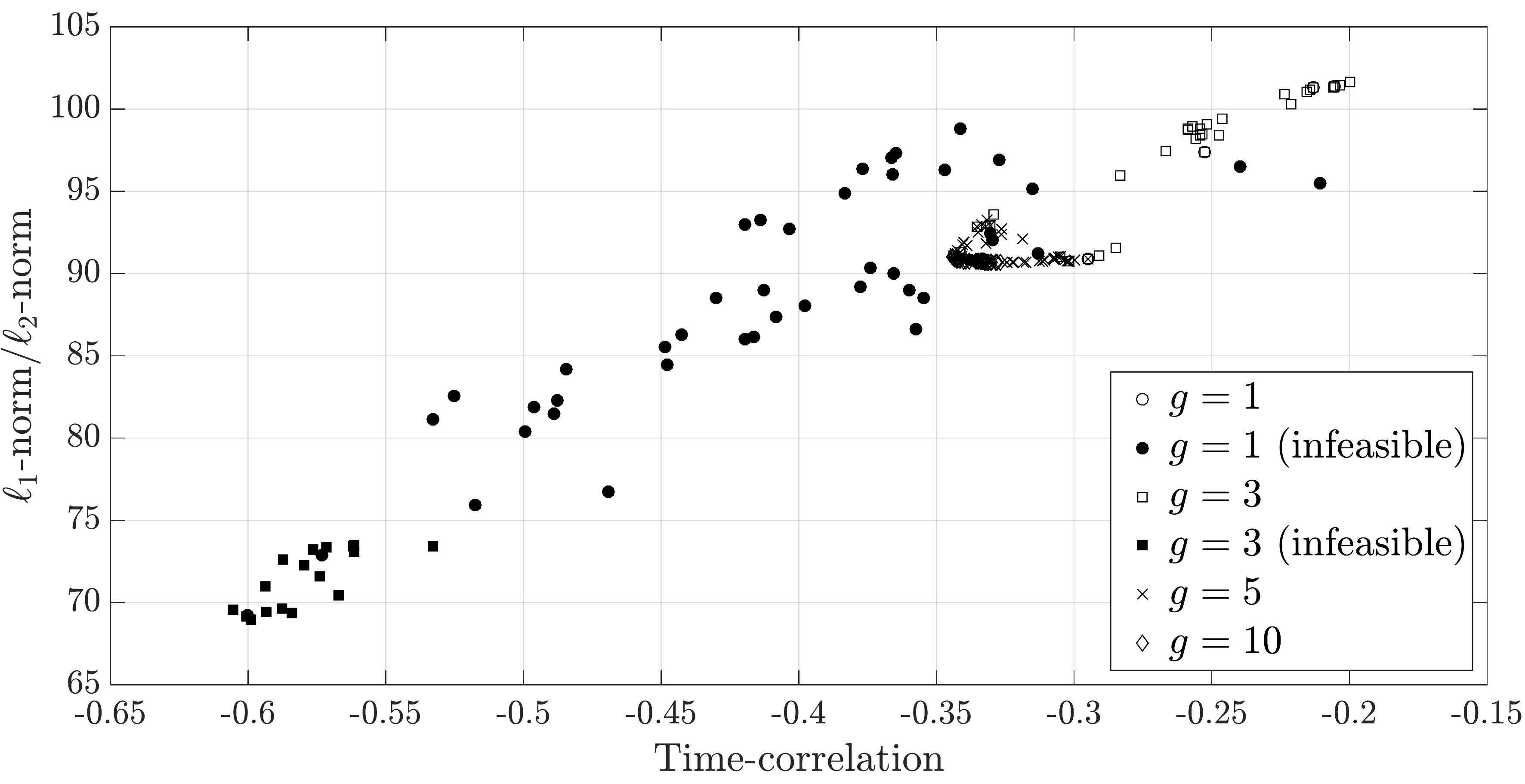}
\label{fig:conv_first}}
\hfil
\subfloat[Last iterations.]{\includegraphics[width=5.5in]{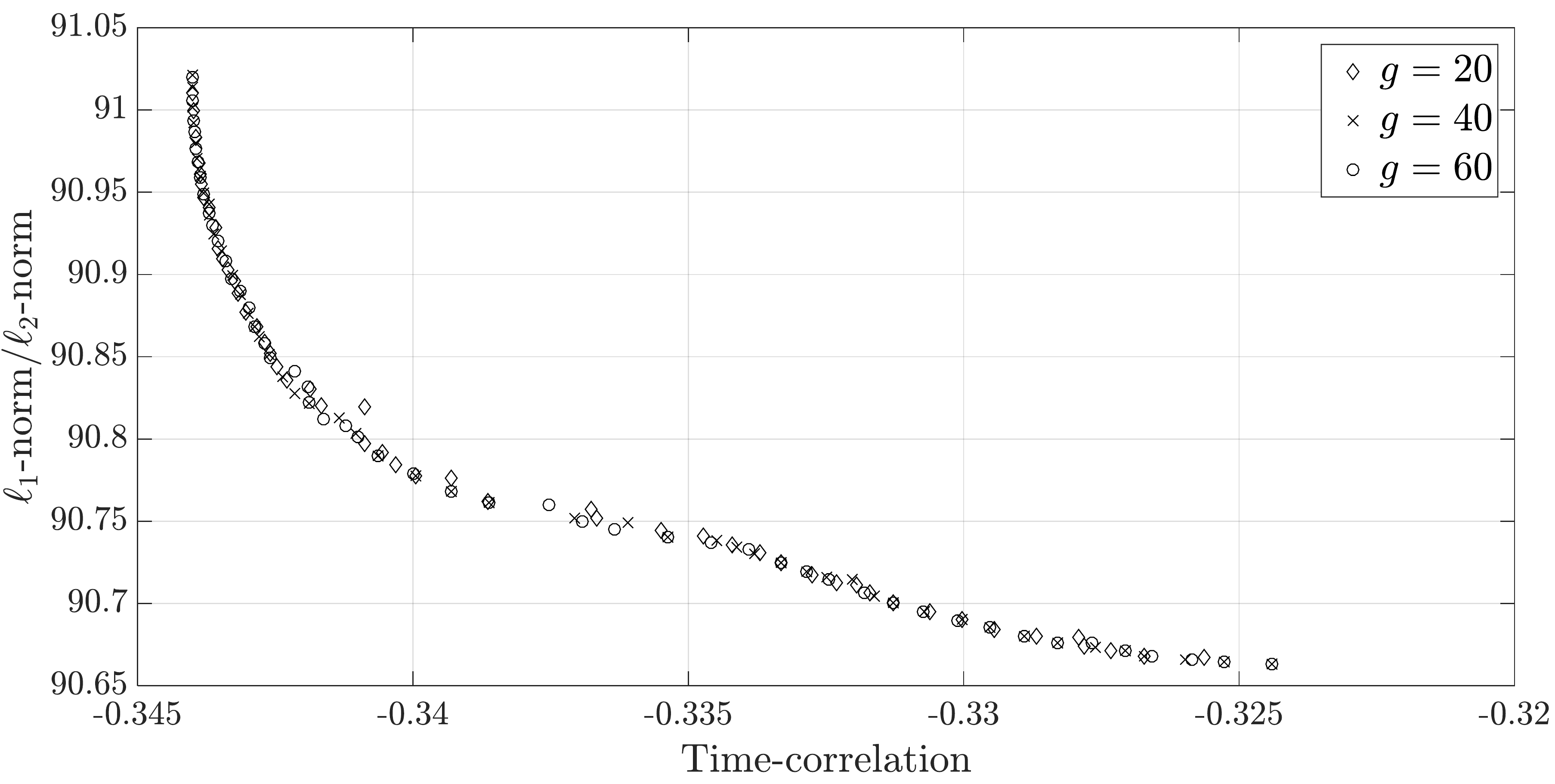}
\label{fig:conv_last}}
\caption{Convergence of SPEA2 algorithm.}
\label{fig:conv_spea2}
\end{figure}

Aiming at investigating the convergence of SPEA2, we provide in Figure~\ref{fig:conv_spea2} the non-dominated solutions (stored in the external set) achieved through the iterations of the algorithm.

Results in Figure~\ref{fig:conv_first} indicates that, in the beginning of the algorithm, one has several infeasible individuals. However, after some iterations, one eliminates them and keep only the feasible ones. Moreover, one may note that, in the first iterations, the SPEA2 promotes a fast convergence toward the non-dominated set. On the other hand, in the last iterations (illustrated in Figure~\ref{fig:conv_last}), the algorithm works more on the diversity of this set. It worth to remark that, in this case, the considered maximum number of iterations ($G = 60$) was sufficient to achieve convergence and diversity of the non-dominated solutions.

In terms of computational complexity\footnote{Computing device: Intel Core i7, 2.20 GHz, 8.00 GB RAM, software MATLAB 2015.}, since we have several iterations of SPEA2, the MO-BSS approach demanded more time compared to the mono-objective formulation. For instance, the computational time of the mono-objective optimization for the time-correlation and the sparsity criteria were 0.4577 and 0.6552 seconds, respectively. For the MO-BSS approach, the execution took 9.1980 seconds (or 0.1533 seconds for each iteration), which is a feasible time in a large number of applications.

\subsection{Experiment with a noisy mixture}

Differently from the experiment conducted in Section~\ref{subsec:explin}, we consider here the same mixing process described in~\eqref{eq:mixture}, with additive white Gaussian noise (AWGN) according to the signal-to-noise ratio (SNR), given by:
\begin{equation}
SNR = 10 \log \frac{\sigma_{signal}^2}{\sigma_{noise}^2},
\end{equation}
where $\sigma_{signal}^2$ and $\sigma_{noise}^2$ are the powers of the signal and the noise, respectively.

\begin{figure}[ht]
\centering
\includegraphics[height=7.0cm]{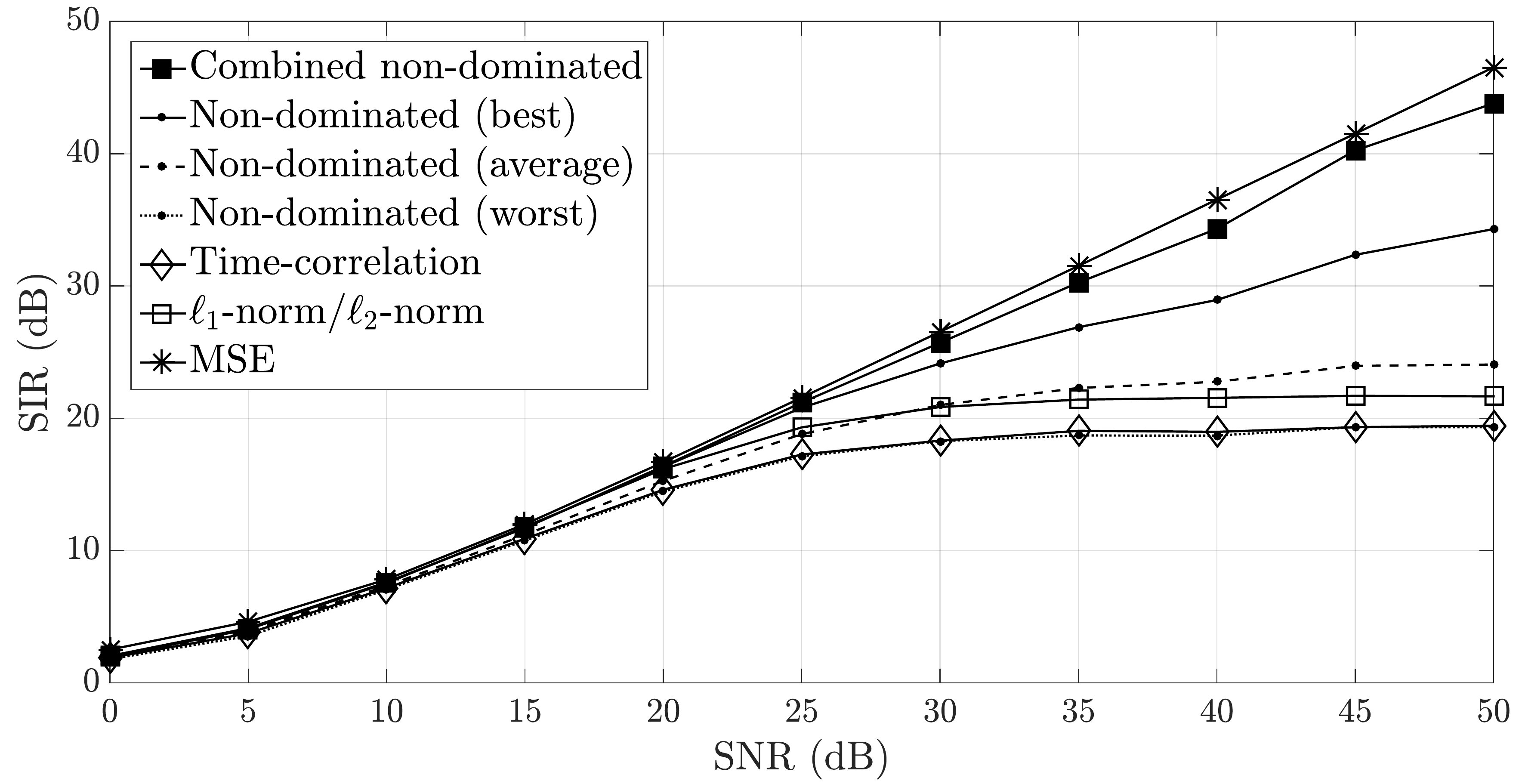}
\caption{SIR values for different AWGN.}
\label{fig:bsssir}
\end{figure}

For each SNR in the range $(0,50]$ (in dB), we apply the SPEA2 algorithm and calculate the SIR value for the solutions obtained by optimizing each criterion individually, the best, the worst and the average among the non-dominated solutions, the MSE solution and the combined non-dominated solution. Based on 100 simulations for each SNR value, the mean values ​​of SIR are shown in Figure~\ref{fig:bsssir}.

In the presence of noise, the results shown in Figure~\ref{fig:bsssir} also indicate that one achieves better solutions in the non-dominated set, based on the SIR values, compared to the ones obtained from the optimization of each criterion individually. This is verified, specially, for SNR greater than 20 dB. Moreover, the combined non-dominated solution achieves SIR values really close to the MSE solution.

\section{Conclusions}
\label{sec:conc}

In many real situations modeled as blind source separation problems, one does not have a separation criterion that can lead to a perfect separation of the source signals. However, one may have a set of prior information that can be used to formulate different separation criteria in order to achieve a better estimation of the sources. In this paper, we proposed to use this set of separation criteria in a multi-objective optimization approach to deal with BSS problems.

Since most of the existing BSS methods found in the literature deal with the addressed problem by taking into account a single characteristic about the source signals, the output of the system comprises a single estimation of these signals. Even for the cases in which the separation method takes into account more properties, the problem formulation relies on the aggregation of several separation criteria into a single one, which, in turn, leads to mono-objective optimization problem whose solution is a single estimation of the source signals. Therefore, the obtained result is dependent on an \textit{a priori} decision about the aggregation procedure, since the user must define (previously to the adaptation) a set of weighting factors for each separation criterion.

On the other hand, our proposal takes into account a set of characteristics about the signal sources and optimize the associated separation criteria simultaneously. As a consequence, one obtains a set of estimations, which can provide more information for the system user take his/her decision. However, in order to apply the multi-objective formulation, one needs to utilize an optimization method that can handle a vector of cost functions, such as the ones based on heuristics. Therefore, a price to be paid is an increase in terms of computational complexity, since the mono-objective methods tend to be faster compared to the multi-objective ones.

The results obtained in our numerical experiments attested the applicability of a multi-objective formulation in the addressed BSS problem. The non-dominated set found by the MO-BSS technique includes a better solution (in terms of the SIR value) compared to the ones obtained from the optimization of a single criterion. With respect to the selection of a proper non-dominated solution, since the problem is blind, it is not possible to know the one that maximizes the SIR value. However, as one could visualize in the obtained results, practically any solution within the non-dominated set is better compared to the ones achieved by the mono-objective formulation. Therefore, a non-dominated solution selected by the system user tends to be a good one. Moreover, he/she can also combines estimates from different solutions in the non-dominated set in order to achieve even better estimates.

As a final remark, the contribution of this paper lies in a separating system that is few exploited in the literature of blind source separation. The existing expert systems often provide a single output for the problem, however, a set of estimates is suitable in several applications. For instance, a set of signals or images could be useful for a physicians to recognize a disease or to indicate a proper treatment. Therefore, since our proposal is general and can be easily adapted to take into account other separation criteria, it becomes possible to consider the use of the multi-objective formulation in others BSS problems.

There are several perspectives to be considered. A first one is to consider different separation criteria so information of different nature can be also be taken into account in the solution. A second perspective, which is somehow connected with the first one, is how to tackle the situations in which there is a great number of separation criteria and also a larger number of sources and mixtures. In these cases, the resulting optimization problem may be tricky.

There are also perspectives related to the mixing models. In many applications~\citep{Comon2010}, the mixtures stem from nonlinear models. As a consequence, the separation criteria may lead to an objective space in which nonlinearities are strongly present.  

Finally, as a future work, we plan to apply multicriteria decision analysis (MCDA)~\citep{Figueira2005}, which have been extensively used in the development of expert systems~\citep{Behzadian2012,Chai2013,Zyoud2017}, in order to select a subset from the non-dominated set. \\

\noindent \textbf{Acknowledgments} \\

The authors would like to thank FAPESP (Processes n. 2014/27108-9 and 2015/16325-1) and CNPq (Processes n. 311786/2014-6 n. 305621/2015-7) for the financial support. \\



\bibliography{_references}

\end{document}